# Etching of elemental layers in oxide molecular beam epitaxy by O$_2$-assisted formation and evaporation of their volatile suboxide: The examples of Ga and Ge

Wenshan Chen, Kingsley Egbo, Huaide Zhang, Andrea Ardenghi, and Oliver Bierwagen

Paul-Drude-Institut für Festkörperelektronik, Leibniz-Institut im Forschungsverbund Berlin e.V., Hausvogteiplatz 5–7, 10117 Berlin, Germany

## Abstract

The delivery of an elemental cation flux to the substrate surface in the oxide molecular beam epitaxy (MBE) chamber has been utilized not only for the epitaxial growth of oxide thin films in the presence of oxygen but also in the absence of oxygen for the growth temperature calibration (by determining the adsorption temperature of the elements) and *in-situ* etching of oxide layers (e. g., Ga$_2$O$_3$ etched by Ga). These elemental fluxes may, however, leave unwanted cation adsorbates or droplets on the surface, which traditionally require removal by *in-situ* superheating or *ex-situ* wet-chemical etching with potentially surface-degrading effects. This study demonstrates a universal *in-situ* approach to remove the residual cation elements from the surface via conversion into a volatile suboxide by a molecular O$_2$-flux in an MBE system followed by suboxide evaporation at temperatures significantly below the elemental evaporation temperature. We experimentally investigate the *in-situ* etching of Ga and Ge cation layers and their etching efficiency using *in-situ* line-of-sight quadrupole mass spectrometry (QMS) and reflection high-energy electron diffraction (RHEED). The application of this process is demonstrated by the *in-situ* removal of residual Ga droplets from a SiO$_2$ mask after structuring a Ga$_2$O$_3$ layer by *in-situ* Ga-etching. This approach can be generally applied in MBE and MOCVD to remove residual elements with vapor pressure lower than that of their suboxides, such as B, In, La, Si, Sn, Sb, Mo, Nb, Ru, Ta, V, and W.

## Introduction

Transparent semiconducting oxides like Ga$_2$O$_3$, In$_2$O$_3$, SnO$_2$, and GeO$_2$ have been rediscovered as promising (ultra-)wide band gap semiconductors for applications in power electronics.[1–5] Their growth as epitaxial thin films by molecular beam epitaxy (MBE) is beneficial for materials exploration and device applications, both requiring a high-degree of purity and crystallinity.

In the MBE growth of oxides possessing volatile suboxide such as Ga$_2$O$_3$, In$_2$O$_3$, SnO$_2$, and GeO$_2$ from elemental sources (Ga, In, Sn, and Ge), the provided elemental flux is oxidized via a first oxidization step to form suboxide (Ga$_2$O, In$_2$O, SnO, and GeO) on the substrate. The suboxide is further oxidized via a second oxidization step to form solid oxide thin film.[6,7] The competing desorption of the intermediately formed suboxide (typically having a higher vapor pressure than its cation element) can decrease the thin film growth rate. While in an oxide MBE-growth chamber, suboxides were also found to form (and evaporate) readily from the elemental sources at a typical molecular $O_2$ background pressure present during growth[8,9], their oxidation into the stable oxide (e.g. $Ga_2O_3$, $In_2O_3$, $SnO_2$, $GeO_2$ and $SiO_2$) required more reactive oxygen species, e.g. provided by an oxygen plasma.[7,8,10–13] Solely, the growth of $In_2O_3$ at a low growth rate of 0.6 nm/min has been demonstrated using molecular $O_2$.[14]

Beyond the mere epitaxy, delivering an elemental cation flux to a substrate surface at absent anion flux in the vacuum of the MBE growth chamber has been used for substrate-temperature calibration purposes[15] or as an *in-situ* oxide removal technique to remove the native $Ga_2O_3$ from GaAs (or GaN) substrates by delivering a Ga flux ("Ga polishing")[16,17]. In the oxide-removal process, the provided element reacts with the oxide into a volatile suboxide, e.g.,

$$4Ga + Ga_2O_3 \rightarrow 3Ga_2O \tag{1}$$

or

$$Ge + GeO_2 \rightarrow 2GeO \tag{2}$$

which desorbs at elevated substrate temperature.[7,18] The *In-situ* oxide removal is beneficial not only for preparing a clean substrate surface prior to growth but can speed up the MBE growth routine by regaining a fresh substrate surface after in-situ growth calibration or unsuccessful oxide layer growth, thus eliminating the need for unloading/loading of substrates and associated temperature ramps for each growth attempt. Meanwhile, it has even been used as damage-free etching to structure highly scaled vertical and lateral 3D $Ga_2O_3$-based devices. [19]

Despite these beneficial applications, the elemental fluxes are prone to leave unwanted elemental adsorbates, layers, or droplets on the surface. The removal of these elemental layers, requires heating to the desorption temperature of the element, high-energy sputtering, or *ex-situ* wet-chemical etching — all of which may create unacceptable degradation of the surface.

This work demonstrates a universal *in-situ* approach to remove the respective elemental layer from a substrate surface by exposure to molecular $O_2$. The technique consists of heating the elemental layer to the desorption temperature of its volatile suboxide (typically well below that of the cation element), exposing it to $O_2$ to induce suboxide formation, e.g.,

$$4Ga + O_2 \rightarrow 2Ga_2O \tag{3}$$

or

$$2Ge + O_2 \rightarrow 2GeO \tag{4}$$

followed by suboxide desorption[20–22] as schematically shown in Fig. 1. We investigated the *in-situ* etching of Ga and Ge cation layer by an $O_2$-flux experimentally on 2-inch $Al_2O_3$(0001) substrates in an MBE system and studied the etching efficiency (suboxide-flux/$O_2$-flux). Our results indicate successful etching of Ga and Ge where ≈2.1% and ≈1.8% of the provided $O_2$ contributed to their removal. Finally, we demonstrate the application scenario of Ga-droplet removal from a $SiO_2$ mask after *in-situ* structuring of a $Ga_2O_3$ layer by Ga-etching.

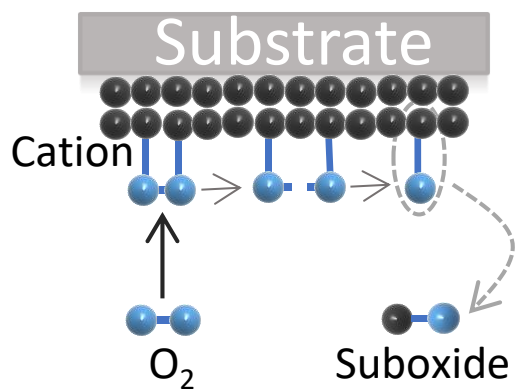

Fig. 1. Schematic describing the *in-situ* etching of cation layer at elevated substrate temperature by an O$_2$-flux, including the physisorption, dissociation, and chemisorption of O$_2$, followed by the desorption of the formed volatile suboxide.

## Experimental details

For this study, Ge and Ga cation layers were grown in high vacuum (background pressure 10$^{-8}$ mbar) on 2-inch c-plane sapphire (Al$_2$O$_3$(0001)) wafers at a temperature of 400 °C by MBE. The rough backside of the single-side polished substrate was sputter-coated with titanium to allow for non-contact substrate heating by radiation from the substrate heater. The substrate temperature ($T_{sub}$) was measured with a thermocouple placed behind the substrate heater. Standard shuttered effusion cells were used to evaporate Ge (7N purity) and Ga (7N purity) from pyrolytic BN crucibles. The beam equivalent pressure (BEP) of the cations and O$_2$, proportional to the particle flux, were measured by a nude filament ion gauge positioned at the substrate location. The BEPs are given in units of mbar and are converted into the equivalent particle flux (atoms cm$^{-2}$s$^{-1}$) by multiplying the measured growth rate of the GeO$_2$ and Ga$_2$O$_3$ layer under conditions of full Ge and Ga incorporation by the cation number density of Ge (4.6 × 10$^{22}$ cm$^{-3}$) and Ga (4.4 × 10$^{22}$ cm$^{-3}$) and using kinetic gas theory in the case of O$_2$. For the layer deposition, the used Ge and Ga -cell temperatures of 1300 °C[1] and 900 °C resulted in Ge and Ga-fluxes ($\Phi$) of $\Phi_{Ge}$ = 4.6 × 10$^{14}$ cm$^{-2}$s$^{-1}$ and $\Phi_{Ga}$ = 1.35 × 10$^{14}$ cm$^{-2}$s$^{-1}$ impinging on the substrate.

Next, we provided molecular O$_2$ to etch the deposited Ga and Ge layers at elevated substrate temperatures that allow the forming suboxides to desorb. For this purpose, a mass flow controller supplied molecular O$_2$ from the research-grade O$_2$ gas (6N purity) and the O$_2$ flow was set as standard cubic centimeters per minute (sccm). The flux $\Phi$ of desorbing species from the layer surface was measured *in-situ* by line-of-sight quadrupole mass spectrometry (QMS, Hiden Analytical "HAL 511 3F"). The QMS ionizer was run at an electron energy of 50 eV to obtain optimal sensitivity. Therefore, some of the measured signals might be affected by fragmentation of suboxide molecules into cation and oxygen atoms.[23] To assess the surface coverage, the process was additionally *in-situ* monitored by reflection high-energy electron diffraction (RHEED).

As an application example, we demonstrated the *in-situ* removal of Ga-droplets from a SiO$_2$ mask directly after *in-situ* patterning of Ga$_2$O$_3$ by Ga-etching. For this purpose, an MBE-grown,

---

[1] Note, that this high temperature lead to a relatively fast degradation of the used standard effusion cell.

≈500 nm-thick $Ga_2O_3$ layer was covered by a ≈75 nm-thick $SiO_2$ hard mask (deposited using sputtering and structured by contact lithography and $CF_4$-based reactive ion etching) and subsequently loaded into the MBE growth chamber. Ga-etching was performed by exposure to a Ga flux of $\Phi_{Ga}$ = 7.8 × $10^{14}$ $cm^{-2}s^{-1}$ at $T_{sub}$ =650°C in the absence of $O_2$ for a total etching time of 40 min, resulting in an etch depth of ≈140 nm (determined by profilometry measurement). Subsequently, we *in-situ* removed the Ga droplets that remained on the $SiO_2$ mask by exposure to 1 sccm $O_2$ for 90 min at the same $T_{sub}$. The untouched structured-$Ga_2O_3$ sample, the structured -$Ga_2O_3$ sample after Ga-etching, as well as a Ga-etched $Ga_2O_3$ sample after molecular $O_2$ exposure were observed by top-view scanning electron microscopy (SEM).

## Results and Discussion

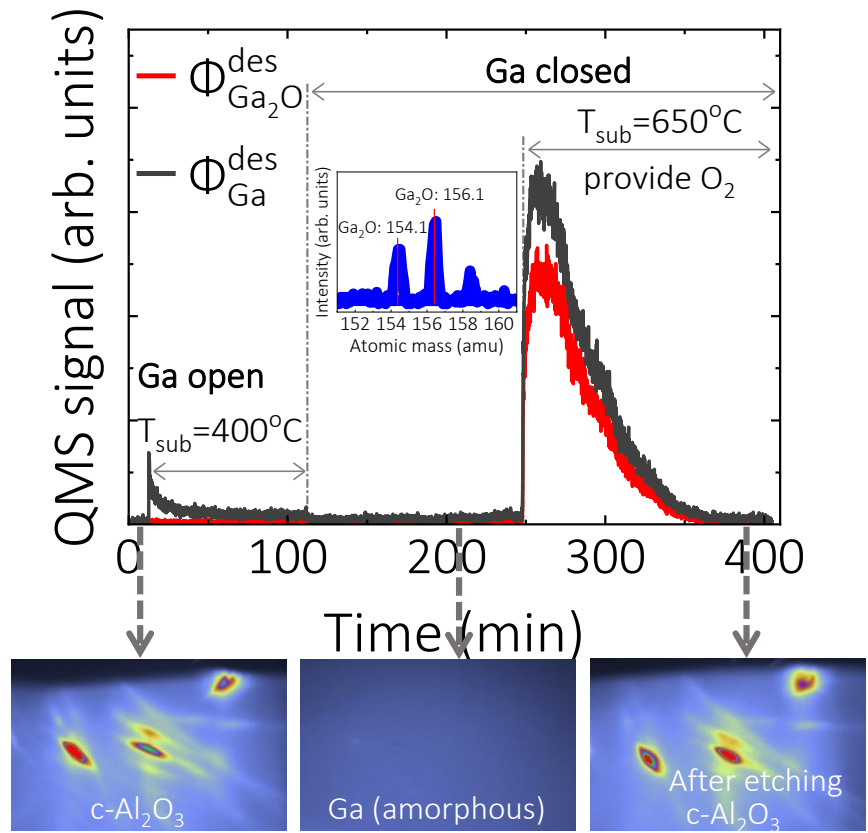

Fig. 2. Ga deposition and its $O_2$-assisted removal. The upper figure shows the measurement of the desorbing flux of $^{69.1}$Ga and $^{156.1}Ga_2O$ by QMS. Three stages are depicted: the deposition of the Ga layer on the c-plane sapphire substrate, the increase of the substrate temperature ($T_{sub}$) to enable suboxide desorption, and the subsequent *in-situ* etching of the already grown Ga layer. The corresponding Ga shutter opening and closing, $T_{sub}$ as well as period of $O_2$ supply are marked. The inset shows the mass spectrum of $Ga_2O$ detected by QMS. The arrows point to images of the RHEED pattern during different stages of the experiments.

Fig. 2 presents the QMS signal of Ga and $Ga_2O$ (proportional to the desorbing flux of $\Phi_{Ga}$ and $\Phi_{Ga_2O}$) and the corresponding RHEED images at different stages during the elemental layer deposition and its subsequent etching by $O_2$. When the Ga shutter was opened, the elemental desorption slightly increased and then rapidly faded, corresponding to an almost full adsorption of the provided flux. After closing the Ga shutter, $T_{sub}$ was immediately increased to 650 °C at 0.5 °C/s to facilitate the $Ga_2O$ desorption in the following etching process. The elevated $T_{sub}$ did not result in detectable desorption of the already grown cation layers (the QMS signal before supplying $O_2$ is negligible), while the disappeared streaky RHEED pattern (middle) clearly indicates a substrate coverage by this layer. A dramatic $\Phi_{Ga_2O}$ signal increase can be observed when an $O_2$ flow (1 sccm) started impinging on the surface. This observation confirms, that $O_2$ reacted with Ga to form $Ga_2O$ via Eq. (3) at a temperature that allows the suboxides to desorb. The $\Phi_{Ga_2O}$ fades gradually from the maximum value, likely due to the gradual decrease of surface fraction covered by the elemental layer. The Ga signal during etching is related to the fragmentation of $Ga_2O$ molecules by the electrons of the ionizer in the quadrupole mass spectrometer. The complete removal of the Ga layer is evidenced by the disappearance of the $\Phi_{Ga_2O}$ signal and by the reappearance of the streaky RHEED pattern of the substrate.

To determine the efficiency of the etching process, we established a quantitative relation of impinging $O_2$ flux and desorbing $Ga_2O$ flux at varying flow rates of $O_2$. Fig. 3(a) illustrates the QMS signal of $\Phi_{Ga_2O}$ during the deposition of 6 equal layers of metallic Ga and their *in-situ* etching by $O_2$ at a decreasing flow, which were 2.00, 1.50, 1.00, 0.80, 0.50, 0.25 sccm, respectively. These experiments were carried out in sequence using Ga deposition and etching temperatures of 400 °C and 650 °C, respectively. Similar to Fig. 2, a sharp increase of $\Phi_{Ga_2O}$ was detected when $O_2$ was supplied, and different $O_2$ fluxes were able to fully convert the Ga layers into evaporated $Ga_2O$, leaving behind a clean surface. Apparently, the maximum $\Phi_{Ga_2O}$ decreases with reduced impinging $O_2$ flux and the required time to completely remove the same amount of Ga increases simultaneously.

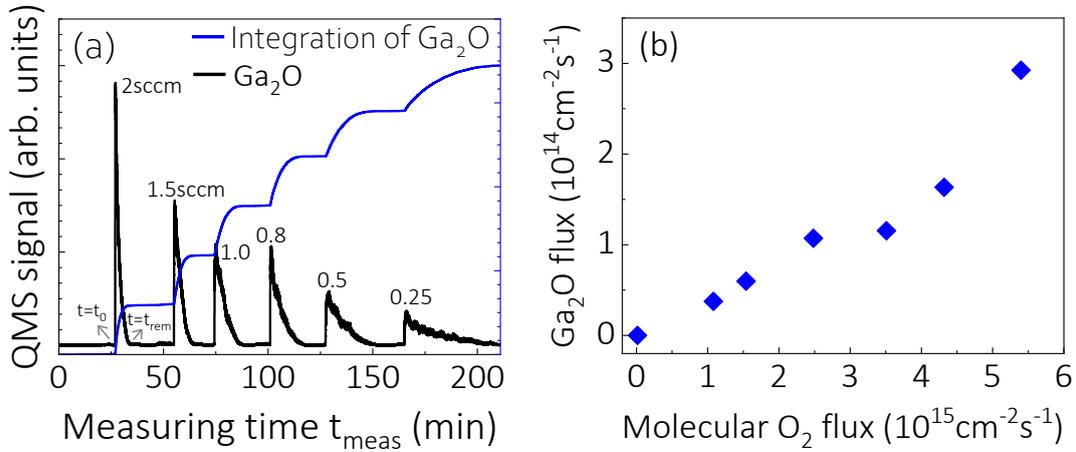

Fig. 3. Ga-deposition/$O_2$-assisted etching cycles using six different decreasing $O_2$ fluxes. (a) Detected $Ga_2O$ flux by QMS as a function of time $t_{meas}$ during metallic Ga layer deposition and its $O_2$-assisted etching. (b) Calibrated desorbing $Ga_2O$ flux during etching as a function of provided $O_2$ flux in Fig. 3(a).

Next, we quantified and related the $\Phi_{Ga_2O}$ measured by QMS and the impinging $O_2$ flux, as presented in Fig. 3(b). All Ga layers were deposited with a fixed Ga flux of $1.35 \times 10^{14}$ cm$^{-2}$s$^{-1}$ over a period of 780 s, resulting in a total surface Ga-atom coverage of $D_{Ga} = 1.05 \times 10^{17}$ cm$^{-2}$. By numerically integrating the QMS signal $Q(t)$ for $Ga_2O$ from the time when $O_2$ is supplied ($t = t_0$) to the time when the whole layer is removed ($t = t_{rem}$), an equivalence relationship of

$$D_{Ga}/2 = \alpha \int_{t=t_o}^{t=t_{rem}} Q(t)dt \tag{4}$$

was obtained and allowed us to determine the calibration factor $\alpha$ that converts the QMS-signal $Q$ (arb. units) into the desorbing molecular flux $\Phi_{Ga_2O}$ ($Ga_2O$ cm$^{-2}$s$^{-1}$). To determine the fraction of the provided O$_2$ species that can contribute to the removal of Ga, the resulting O$_2$ flux $\Phi_{O_2}$ used at the different O$_2$ flow rates is calculated based on kinetic gas theory[24] from the corresponding measured O$_2$-BEP ($P_{O_2}$) according to

$$\Phi_{O_2} = P_{O_2} \times (N_A/2\pi M K_B T)^{1/2} \quad (5)$$

with the Avogadro constant $N_A$, the molar mass M of O$_2$, and O$_2$ temperature $T$ (298 K). The measured $P_{O_2}$ as a function of O$_2$ flow can be seen in Fig. S1 in supplementary material. Fig. 3(b) shows the peak $\Phi_{Ga_2O} = \alpha Q(t)_{max}$ observed at the beginning of each etching-cycle as a function of the corresponding $\Phi_{O_2}$. Based on Eq. (3), an average etching efficiency ($\eta$) of $\eta = \frac{\Phi_{Ga2O}}{2*\Phi_{O2}} \times 100\%$ = 2.1% was obtained.

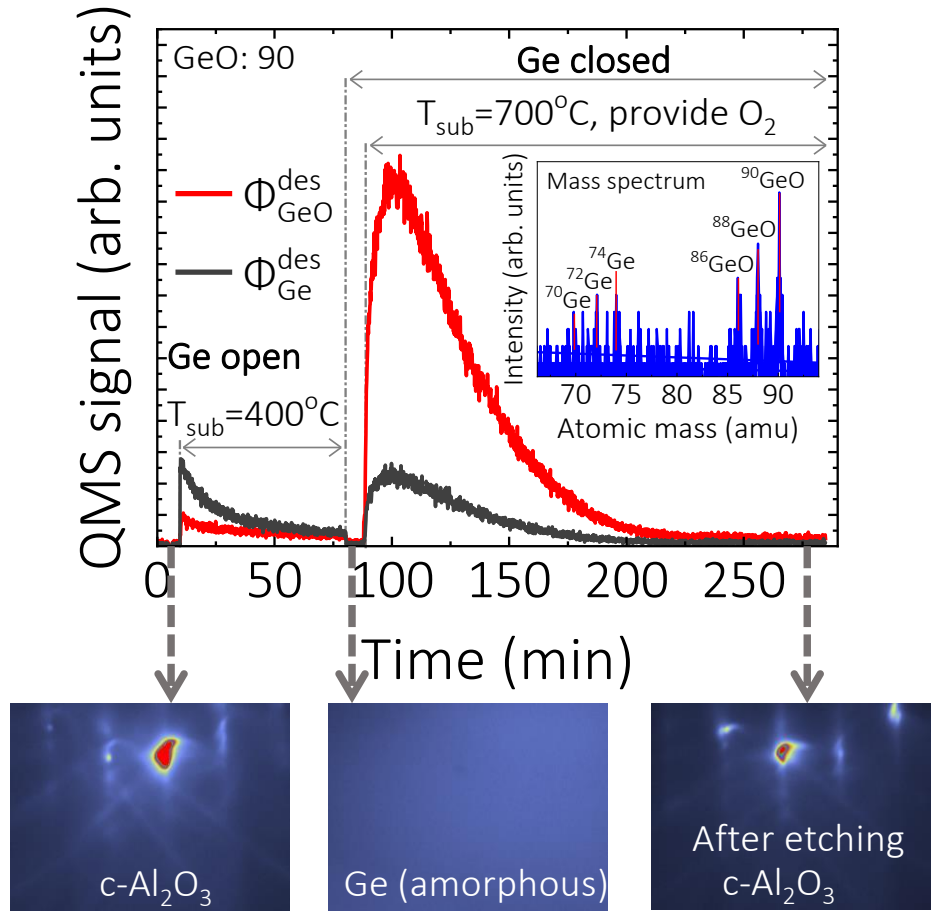

Fig. 4. Ge deposition and its O$_2$-assisted removal. The upper figure shows the measurement of the desorbing flux of $^{74}$Ge and $^{90}$GeO by QMS. Three stages are depicted: the deposition of the Ge layer on the c-plane sapphire substrate, the increase of the $T_{sub}$ to enable suboxide desorption, and the subsequent *in-situ* etching of the already grown Ge layer. The corresponding Ge shutter opening and closing, $T_{sub}$ as well as period of O$_2$ supply are marked. The inset shows the mass spectrum of Ge and GeO detected by QMS. The arrows pointed images show the evolution of the RHEED pattern during different stages of the experiments.

The QMS signal of the desorbing $\Phi_{Ge}$ and $\Phi_{GeO}$, as well as the surface development monitored by RHEED during Ge layer deposition and O$_2$-etching experiment, shown in Fig. 4, exhibit qualitatively similar behavior to that observed for Ga. To enhance the GeO desorption in etching process, O$_2$ was supplied at $T_{sub}$ = 700 °C. A significant $\Phi_{GeO}$ signal increase can be observed when the O$_2$ (1sccm) approached the surface, confirming that O$_2$ reacted with Ge to form GeO via Eq. (4). The disappearance of $\Phi_{GeO}$ and reappearance of the streaky RHEED pattern of the substrate proved a complete removal of the Ge layer. Similarly, we determined $\eta$ for Ge etching using the same methodology employed in our Ga experiment. A surface coverage of $D_{Ge} = 1.68 \times 10^{18}$ cm$^{-2}$, a desorbing flux of $\Phi_{GeO}$ = 1.74 x $10^{14}$ $cm^{-2}s^{-1}$ was obtained based on the using experiment parameters, while 1 sccm of $O_2$ corresponding to a $P_{O_2}$=1.55 × $10^{-5}$ mbar, which can be translated in to $O_2$ flux of $\Phi_{O_2}$ = 4.2 x $10^{15}$ $cm^{-2}s^{-1}$. Consequently, a $\eta_{Ge} = \frac{\Phi_{GeO}}{2*\Phi_{O_2}}$ x 100% = 1.8% representing removal efficiency of the $O_2$ was obtained according to Eq. (4).

To better illustrate the application scope of our studies, we conducted experimental tests on the device structuring process by *in-situ* Ga-etching a $SiO_2$-masked $Ga_2O_3$ layer. Fig. 5 showcases the top view SEM images of the masked and structured $Ga_2O_3$ samples. By comparing Fig. 5(a) with Fig. 5(b), as anticipated, we observed Ga droplets remaining on the $SiO_2$ mask after structuring $Ga_2O_3$ by exposing it to the Ga flux. However, these droplets can be completely removed *in-situ* by providing $O_2$ following the oxide etching process, as evidenced by a clean mask surface depicted in Fig. 5(c).

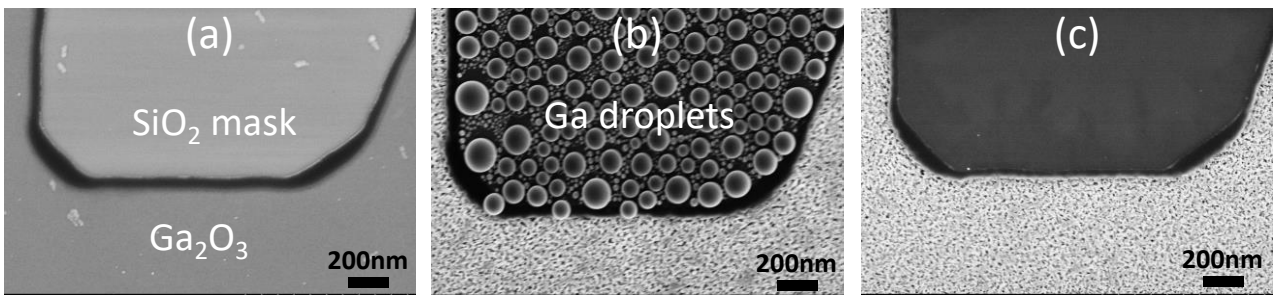

Fig. 5. SEM top view images of (a) the $SiO_2$ mask on $Ga_2O_3$ thin film, (b) the $Ga_2O_3$ thin film etched by a Ga flux with residual Ga droplets on top of the $SiO_2$ mask, and (c) after $O_2$-assisted *in-situ* removal of the Ga droplets from the $SiO_2$ mask.

## Conclusion

In conclusion, we have successfully demonstrated a process in which molecular $O_2$ is utilized to remove elemental Ga and Ge layers in an MBE growth chamber through formation and desorption of their volatile suboxides at temperatures lower than those required for elemental desorption. Under the investigated (non-optimized) conditions about 1.8% - 2% of the provided $O_2$ contributed to cation removal. We showcased the application of this process for the *in-situ* removal of the residual Ga droplets from the $SiO_2$ mask directly after structuring a $Ga_2O_3$ layer by *in-situ* etching using a Ga atomic flux. The $O_2$-assisted cation removal process can be generally applied *in-situ* within an oxide MBE or MOCVD system to remove residual elemental layers that may occur after exposure to the cation fluxes during *in-situ* oxide etching or substrate temperature calibration, and is generally applicable for elemental layers whose suboxide exhibits a higher vapor pressure than the respective elements, such as B, In, La, Si, Sn, Sb, Mo, Nb, Ru, Ta, V, and W.[9,25]

## Acknowledgement

The authors thank Hans-Peter Schönherr, Claudia Hermann, Sander Rauwerdink, and Walid Anders for technical support, Steffen Breuer for discussion, as well as Jingxuan Kang for critically reading the manuscript. This work was performed in the framework of GraFOx, a Leibniz-ScienceCampus partially funded by the Leibniz association. W.C. gratefully acknowledges financial support from the Leibniz association under Grant No. K417/2021.

# Supplementary material to:
# Etching of elemental layers in oxide molecular beam epitaxy by O$_2$-assisted formation and evaporation of their volatile suboxide: The examples of Ga and Ge

Wenshan Chen, Kingsley Egbo, Huaide Zhang, Andrea Ardenghi, and Oliver Bierwagen

Paul-Drude-Institut für Festkörperelektronik, Leibniz-Institut im Forschungsverbund Berlin e.V., Hausvogteiplatz 5–7, 10117 Berlin, Germany

The supplementary material includes the beam equivalent pressure (BEP) of the O$_2$ in the MBE chamber as a function of O$_2$ flow.

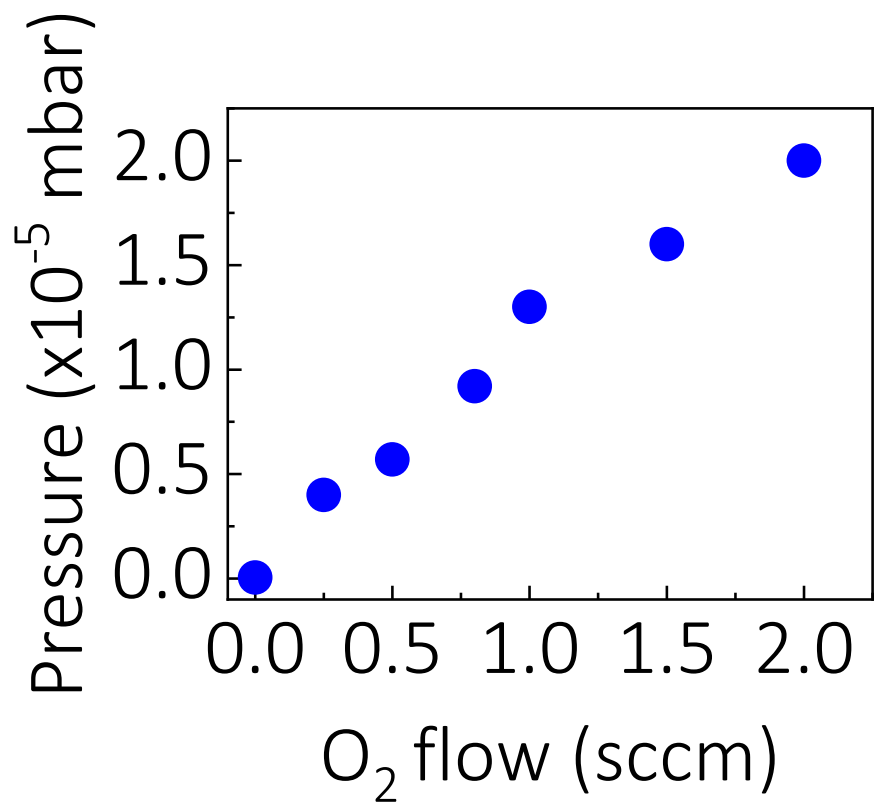

Figure S1: The beam equivalent pressure (BEP) of the O$_2$ in the MBE chamber as a function of O$_2$ flow.